\documentclass{PoS}

\usepackage{amsmath,amssymb}
\usepackage{bbold}
\usepackage{graphicx}
\usepackage{float}
\usepackage [utf8]{inputenc}
\usepackage{subfigure}
\usepackage{euscript}
\usepackage{placeins}

\def\OO{\mathcal{O}}

\def\bx{{\boldsymbol x}}

\def\be{\begin{equation}}
\def\ee{\end{equation}}
\def\cb{c_\mathrm{B}}
\def\gsqa{g_{3\mathrm{d}}^2a}
\def\gsq{g_{3\mathrm{d}}^2}
\def\gfour{g_{3\mathrm{d}}^4}

\def\ycrit{y_{\mathrm{crit}}}
\def\Tr{\mathrm{Tr}}

\title{Full $\mathcal{O}(a)$ improvement of EQCD}

\ShortTitle{Full $\mathcal{O}(a)$ improvement of EQCD}

\author{Guy D. Moore\\
        Institut f\"ur Kernphysik\\
        Schlossgartenstra{\ss}e 2\\
        D-64289 Darmstadt\\
        E-mail: \email{guy.moore@physik.tu-darmstadt.de}}

\author{\speaker{Niels Schlusser}\\
        Institut f\"ur Kernphysik\\
        Schlossgartenstra{\ss}e 2\\
        D-64289 Darmstadt\\
        E-mail: \email{nschlusser@theorie.ikp.physik.tu-darmstadt.de}}

\abstract{EQCD is a 3D bosonic theory containing SU(3) and an adjoint scalar, which
efficiently describes the infrared, nonperturbative sector of hot QCD and which is highly
amenable to lattice study. We improve the matching between lattice and continuum EQCD
by determining the final unknown coefficient in the $\mathcal{O}(a)$ matching, an additive scalar mass
renormalization. We do this numerically by using the symmetry-breaking phase transition
line of EQCD as a line of constant physics. This prepares the ground for a precision study
of the transverse momentum diffusion coefficient $C(q_\perp)$ within this theory. }

\FullConference{37th International Symposium on Lattice Field Theory - Lattice2019\\
		16-22 June 2019\\
		Wuhan, China}

\begin{document}

\section{EQCD as an effective theory of hot QCD}
Quantum Chromodynamics (QCD) is an asymptotically free theory, therefore it should become perturbative at sufficiently high temperatures. However, comparisons of perturbative calculations, for instance the pressure \cite{Kajantie:2002wa} with its nonperturbative lattice results \cite{Bazavov:2014pvz,Borsanyi:2013bia} show a quite big discrepancy even at high temperatures.
The reason for this peculiar behavior lies in the highly occupied gluon-0-mode at high $T$ \cite{Linde:1980ts}. Based on that, one can construct an effective theory that isolates the nonperturbative contribution and integrates out all other degrees of freedom. This theory is called Electrostatic Quantum Chromodynamics (EQCD) and was first proposed by Braaten and Nieto \cite{Braaten:1995cm}. Since it only treats the gluon-0-mode dynamically, it lives in three spatial dimensions. Its continuum action reads
\begin{equation}
S_\mathrm{EQCD,c} =
			\int \mathrm{d}^3x \, \left( \frac{1}{2 \gsq} 
			\mathrm{Tr} \, F^{ij} F^{ij} + 	\mathrm{Tr}  \, D^i \Phi D^i \Phi 
			+ m_\mathrm{D}^2  \mathrm{Tr} \, \Phi^2 + \lambda 
			\big(	\mathrm{Tr}  \, \Phi^2 \big)^2  \right) \, ,
\end{equation}
with the modified gauge coupling $\gsq$, the adjoint scalar field $\Phi$ that is a remnant of the $A^0$-field, its mass $m_\mathrm{D}^2$ and its quartic self-coupling $\lambda$.
The matching calculation that connects the full QCD parameters temperature $T$ and number of flavors $N_\mathrm{f}$ to the effective field theory parameters $\gsq$, $\lambda$ and $m_{3\mathrm{d}}^2$ was carried out in \cite{Kajantie:1997tt,Laine:2005ai} up to at least $\OO(g^4)$.
EQCD lattice simulations have delivered a variety of high-precision results in hot QCD thermodynamics, for instance string tension, correlation lengths and quark number susceptibility \cite{Laine:2005ai,Hart:2000ha,Hietanen:2008xb}.

Furthermore, it has been shown by Caron-Huot that EQCD also describes the interaction of a jet with a hot medium up to subleading accuracy \cite{CaronHuot:2008ni}, which opened the gate for a precise, first-principles prediction of the jet-broadening coefficient $\hat{q}$ from the lattice. Unfortunately, Caron-Huot also found that a purely perturbative computation does not provide converging results and nonperturbative methods are necessary, again.
A first, promising attempt at a lattice calculation for $\hat{q}$ has been made in \cite{Panero:2013pla}. The used lattice action is equivalent to
\begin{align}
S_\mathrm{EQCD,L} =&  \beta \sum_{x,i>j} 
		\left( 1 - \tfrac{1}{3} \square_{ij} \right) 
		+ 2 \sum_{x,i} \mathrm{Tr} \, \left( \Phi_\mathrm{L}^2(x) 
			- \Phi_\mathrm{L}(x) U_i(x) 
			\Phi_\mathrm{L}(x+a\hat{i}) U_i^\dagger(x) \right) \notag \\
			& + \sum_x \Big( Z_2 \left( y + \delta y \right) 
			\mathrm{Tr} 	\, \Phi^2_\mathrm{L}(x) +  
			Z_4 \left( x + \delta x \right) 
			\left( \mathrm{Tr} \, \Phi^2_\mathrm{L}(x) \right)^2 \Big) \, ,
\end{align}
where the lattice versions of the continuum parameters $\beta \equiv 6 / \gsqa$, $x \equiv \lambda / \gsq$ and
$y \equiv m_D^2(\bar\mu=\gsq)/\gfour$ appear. 
A continuum limit was not yet provided in \cite{Panero:2013pla} since EQCD suffers from systematic errors linear in the lattice spacing $a$, stemming from the lattice-continuum-parameter matching. Nearly all $\OO(a)$ errors have been removed analytically \cite{Moore:1997np}. The mass squared parameter $y$, however, receives $\OO(a)$-corrections up to three-loop order in lattice perturbation theory. An analytical determination of this last missing contribution is therefore prohibited by sheer complexity of the calculation.

In this work, we developed a numerical algorithm to determine this last contribution. Briefly summarized, we improve all parameters other than $y$ to $\OO(a)$-accuracy. Linear-in-$a$ behavior in a fit of a line of constant physics must therefore stem from the $y$-renormalization. We utilize the phase transition of EQCD as the line of constant physics. In the physically relevant parameter range, this transition is of first order, which makes the determination of the spot of the phase transition especially hard. The application of standard techniques like multicanonical reweighting is numerically extremely costly in this scenario \cite{Kajantie:1998yc}. Consequently, we developed an alternative method in order to efficiently determine the spot of a first-order phase transition on the lattice.

\section{Method}
EQCD has two phases, in which the $\mathbb{Z}(3)$-symmetry is either intact or spontaneously broken. These are indicated by the order parameter $\Tr \, \Phi^3 = 0$ in the symmetric phase and by $\Tr \, \Phi^3 \neq 0$ in the broken phase respectively. For the physically interesting values of $x$, these two phases are separated from each other by a first-order phase transition \cite{Kajantie:1998yc}.

We developed a new method in order to determine the spot of a first-order phase transition on the lattice precisely. Conventional methods like multicanonical reweighting rely on supporting the entire system to tunnel from one to the other phase. Unfortunately, if tunneling is as heavily suppressed as in the present case, multicanonical reweighting will turn out to be rather ineffective \cite{Kajantie:1998yc}. If we generated a lattice configuration in which both phases coexisted and we wondrously guessed the critical value of $y$ at given self-coupling $x$, we would see that, apart from Brownian motion, no phase would expand at the expense of the other. This leaves us with two major questions we would like to address in the following:
\begin{itemize}
\item How to generate these configurations?

\item How to tune the mass towards its critical value for given $x$?
\end{itemize}

\begin{figure}[thbp!] 
	\subfigure[Establishing coexisting phases by temporarily introducing a space-dependent mass and gradually shrinking its amplitude.]{\includegraphics[scale=0.55]{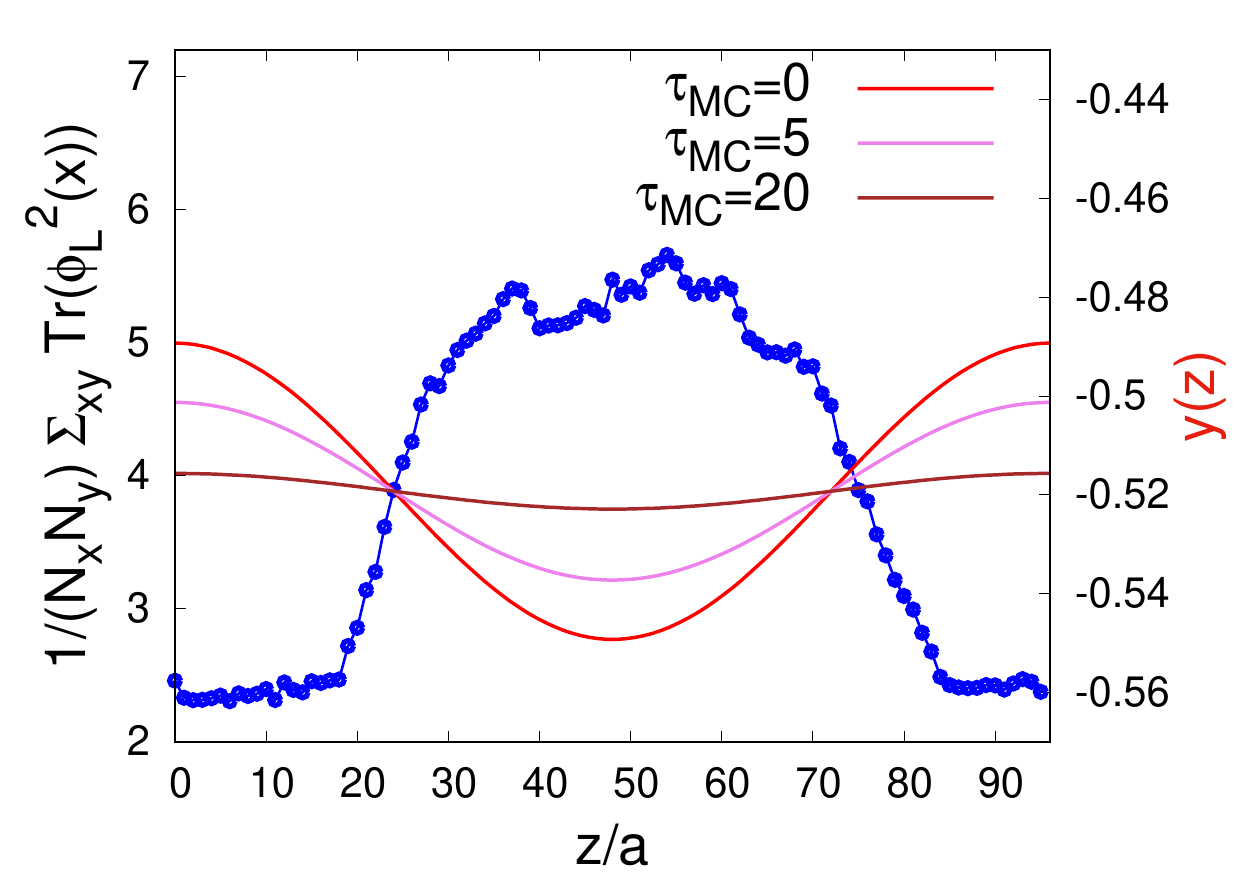}} 
	\subfigure[Tuning the mass towards criticality.]{\includegraphics[scale=0.55]{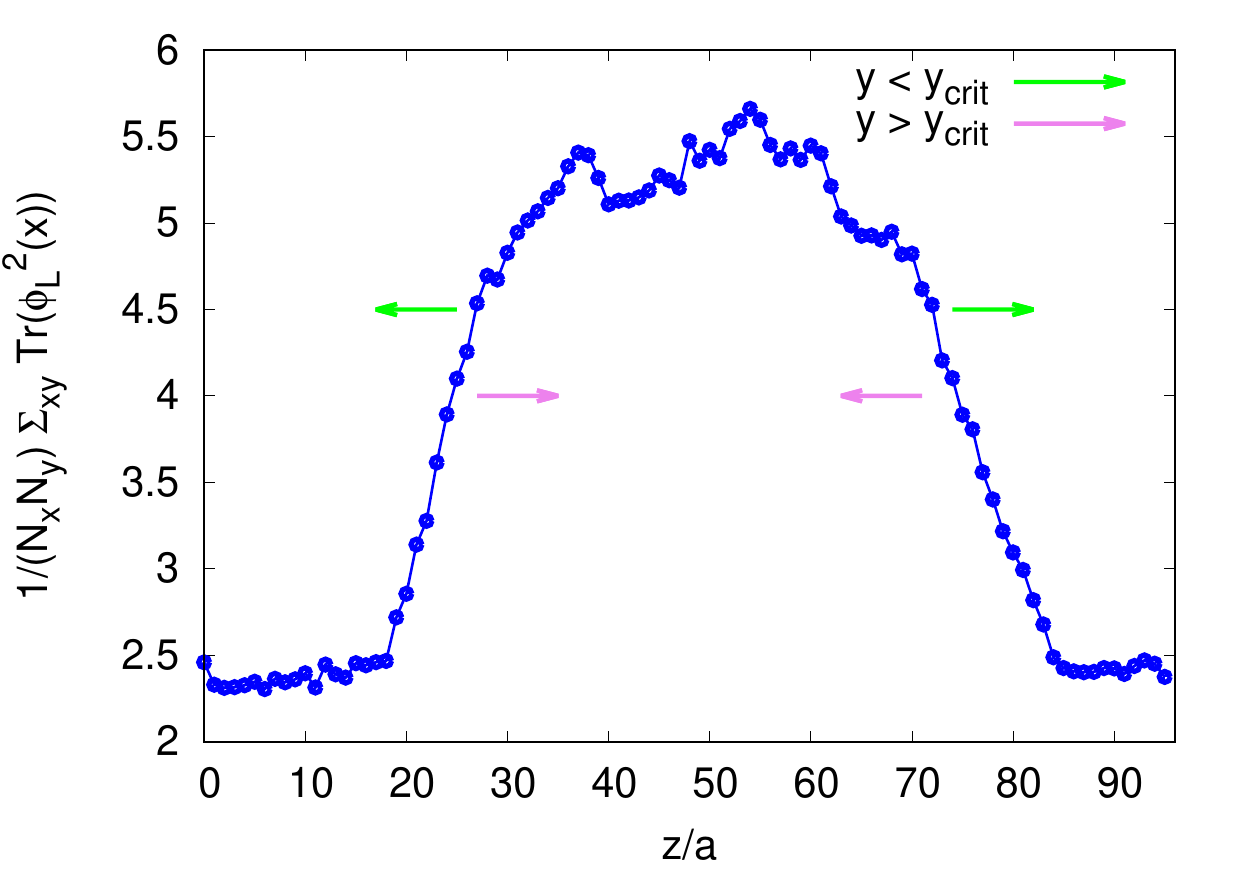}} 
	\label{slices}
	\caption{Two stages of determining $\ycrit$.}
\end{figure}

For the generation of the special configurations, we temporarily introduce a space-dependent mass 
\begin{equation}
 		\label{zdependenty}
  		y(z) = y_{\mathrm{crit,est}} + \Delta y \; \cos(2\pi z/L_z) \, , 
\end{equation}
where $\Delta y$ has to be taken such that the values $y_{\mathrm{crit,est}} \pm \Delta y$ are outside the metastability window in the symmetric (broken) phase. After thermalization, we gradually shrink $\Delta y \to 0$, as depicted in Fig.~\ref{slices}(a). Note that in $\Tr \, \Phi^2$ with averaged transversal directions, the phase transition can also be spotted although it is not the true order parameter. We used it nevertheless since it leads to a more stable phase discriminator.

Already during that process, we have to start tuning $y$ towards its critical value. One possible way to do so is 
\begin{equation}
	\label{mass_adjustment}
	y_{\mathrm{L,\,new}} = y_{\mathrm{L,\,old}} + \cb \cdot \left(
	\frac{\tfrac{1}{V} \sum_{\bx} \Tr \, \Phi^2 - \Tr \,
 	\Phi^2_{\mathrm{symm}}}{\Tr \, \Phi^2_{\mathrm{brok}} - \Tr \, 
 	\Phi^2_{\mathrm{symm}}}  - 0.5 \right) \, , 
\end{equation}
where $c_\mathrm{B}$ is a small coefficient steering the strength of the adjustment, $\Tr \, \Phi^2_\mathrm{brok}$ and $\Tr \, \Phi^2_\mathrm{symm}$ are the values that $\Tr \, \Phi^2$ takes in the particular phase, extracted from separate simulations close to, but not necessarily right on top the critical point. By comparing results at different values of $c_\mathrm{B}$, it was verified that this insertion did not introduce artificial correlations spoiling the result. Note also that although $\Tr \, \Phi^2$ still contains $\OO(a)$-errors, the difference of two $\Tr \, \Phi^2$-operators is free of $\OO(a)$-errors.
If our initial mass $y_{\mathrm{L,\,old}}$ is larger than $\ycrit$, then the symmetric phase is favored and it expands at the expense of the broken phase, as can be seen in Fig.~\ref{slices}(b). According to Eq.~\eqref{mass_adjustment}, this evokes lowering the mass and vice versa for the opposite case. 	

The motion of the phase boundary is driven by a net force of surface area times the free energy difference of the phases $\Delta F$. At small $y-\ycrit$, it is a valid approximation to truncate a polynomial series in $y-\ycrit$ at linear order, leading to agreement of the value where $y$ maintains stability and $\ycrit$. However, if $\mathrm{d}^2 F / \mathrm{d} y^2$, the coefficient of the leading truncation error, becomes large, higher-than-linear orders of the series expansion might significantly bias the result, which we explicitly ruled out for our simulations by doubling $c_\mathrm{B}$ and not observing any significant changes in our results.

\section{Results}
\begin{figure}[thbp!] 
\centering\includegraphics[clip,width=0.9\linewidth]{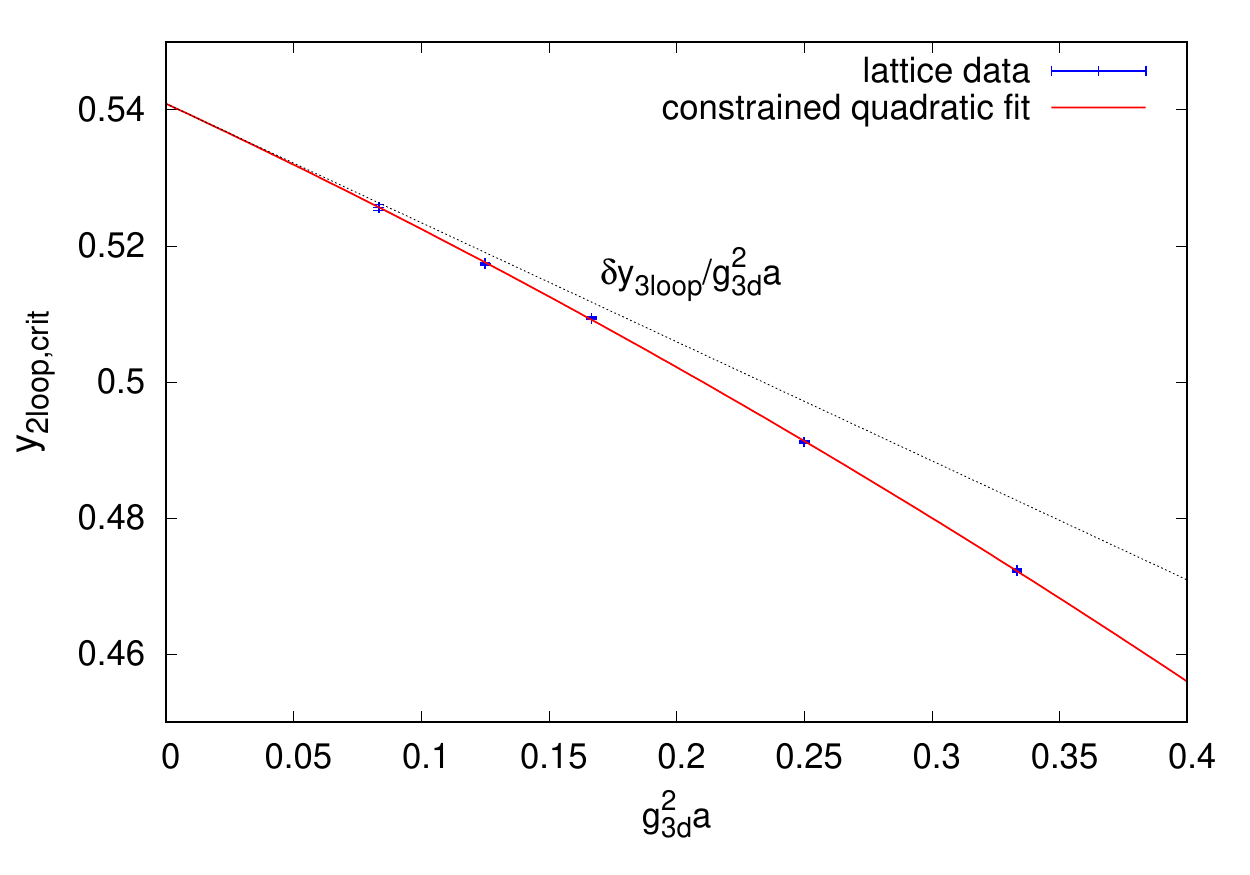}
\caption{Generic example for determination of $\frac{\delta y_{3\mathrm{loop}}}{\gsqa}$, in this case at $x=0.08896$.}
\label{example_x_plot}
\end{figure}

This procedure gives us a tool to efficiently determine $\ycrit$ at given $x$ and lattice spacing $\gsqa$. Repeating this calculation at different lattice spacings $\gsqa$, we can fit the $\OO(a)$-behavior of $\ycrit$ for several values of $x$. This is done by plotting $\ycrit(\gsqa)$ and extracting the counterterm as the slope of a fitted polynomial at $\gsqa=0$, as illustrated in Fig.~\ref{example_x_plot}. In principle, the slope could be measured by a common polynomial fit. However, constrained curve fitting provides a possibility to include our knowledge about the convergence of the perturbative series into that result \cite{Lepage:2001ym}.

\begin{figure}[thbp!] 
\centering\includegraphics[clip,width=0.9\linewidth]{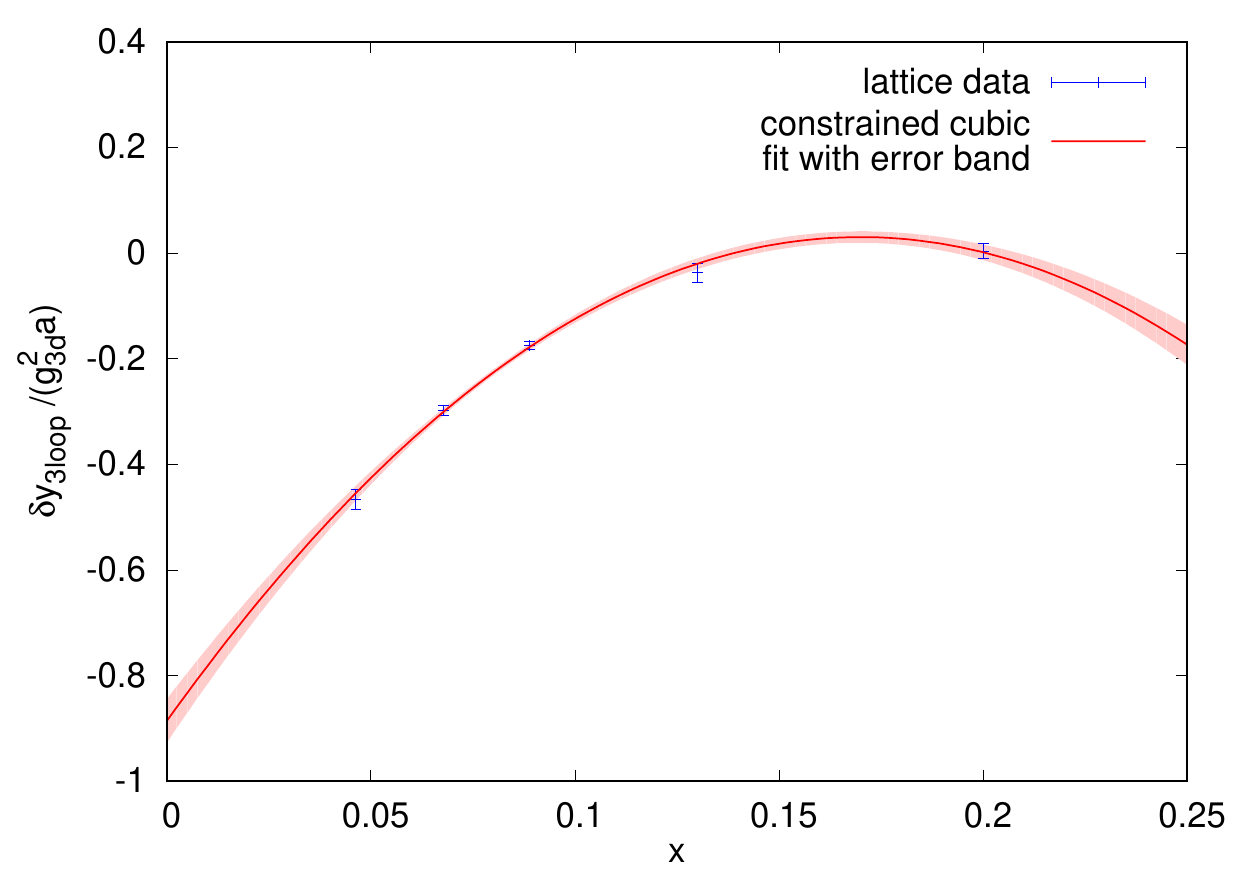}
\caption{Grand fit of $\frac{\delta y_{3\mathrm{loop}}}{\gsqa}(x)$.}
\label{grand_fit}
\end{figure}

We know parametrically that the $\OO(a)$-counterterm has the shape \cite{DOnofrio:2014mld}
\begin{equation}
	\label{form_counterterm}
	\frac{\delta y_{3\mathrm{loop}}}{\gsqa} \, ( x ) = A + B \, x + C \, x^2 + D \, x^3 \, .
\end{equation}
Therefore, determining the counterterm for at least $5$ different values of $x$ allows a grand fit to the given form. Simulation parameters and results can be found in Tab.~\ref{res_coeffs}. The grand fit has been performed in Fig.~\ref{grand_fit}, where we have additionally determined the value of the purely scalar contribution $D$ in a purely scalar computation and included it into the grand fit, again via constrained curve fitting \cite{Lepage:2001ym}. The situation for the purely scalar case is somewhat different because the phase transition is of second order there, which requires a different machinery for the determination of the $\OO(a)$-term. For details, we refer to App.~A in \cite{Moore:2019lua}. The purely scalar coefficient amounts to
\begin{equation}
\frac{m_\mathrm{D}^2}{\lambda^2} = 0.0151(55).
\end{equation}

All self-couplings $x$ in Tab.~\ref{res_coeffs} are in the first-order regime. The first three correspond to physical full QCD scenario at temperatures $T=1\, \mathrm{GeV}, \; 500 \, \mathrm{MeV}, \; 250 \, \mathrm{MeV}$ and numbers of massless fermion flavors $N_\mathrm{f}= 4\,, 3\,, 3$. The last two have no full QCD counterpart but help to constrain the fit over a wider parameter window.

\begin{table}[htbp!]	
\centering
\begin{tabular}{|c|c|c|}	
\hline
$x$ & $y_\mathrm{crit,\,cont}$ & $\delta y_{3\mathrm{loop}} / \gsqa $  \\
\hline
$0.0463596$ & $0.9293(13)$ & $-0.467(19)$ \\
$0.0677528$ & $0.67627(85)$ & $-0.298(10)$ \\
$0.08896$ & $0.54092(76)$ & $-0.1750(74)$ \\
$0.13$ & $0.4043(18)$ & $-0.037(18)$ \\
$0.2$ & $0.2961(15)$ & $0.004(15)$ \\
\hline
\end{tabular}
\caption{Results of our five EQCD simulation sets.}
\label{res_coeffs}
\end{table}

The grand fit in fig.~\ref{res_coeffs} leads to the resulting 3-loop-counterterm
\begin{equation}
  	\label{mainresult}
	\frac{\delta y_{3\mathrm{loop}}}{\gsqa}(x) = 0.0151(55) \,
	x^3 - 31.8(28) \, x^2 + 10.80(74) \, x - 0.886(41) 
\end{equation}
with a plausible value of $\chi^2$.

We see that neither the value nor the error of the cubic coefficient changed. This is a result from the small relative error of the scalar coefficient that is exceeded by all other errors of full EQCD simulations. This also explains why the value of the cubic coefficient was left unchanged by the grand fit. For a precise determination of the errors at any desired $x$, we report the full covariance matrix in Tab.~\ref{cov_mat_grand_fit}. The error band in Fig.~\ref{grand_fit} also requires information from the full covariance matrix.

\begin{table}
\centering
\begin{tabular}{|c|c c c c| }	
\hline
$\mathrm{cov}(C_i,C_j)$ & $C_0$ & $C_1$ & $C_2$ & $C_3$   \\
\hline
$C_0$ & $0.001700$ & $-0.02997$ & $0.1101$ & $-3.563 \cdot 10^{-8}$ \\
$C_1$ & $-0.02997$ & $0.5451$ & $-2.046$ & $1.129 \cdot 10^{-6}$ \\
$C_2$ & $0.1101$ & $-2.046$ & $7.899$ & $-1.079 \cdot 10^{-5}$ \\ 
$C_3$ & $-3.563 \cdot 10^{-8}$ & $1.129 \cdot 10^{-6}$ & $-1.079 \cdot 10^{-5}$ & $3.025 \cdot 10^{-5}$ \\
\hline
\end{tabular}
\caption{Covariance matrix of the grand fit.}
\label{cov_mat_grand_fit}
\end{table}

Further interesting information on the EQCD phase transition, eg.\ on the updated EQCD phase diagram, the strength of the transition and the $\Tr \, \Phi^2$-operator improvement, can be found in \cite{Moore:2019lua}.

\section{Conclusions and Outlook}
In this work, we determined the last missing $\OO(a)$-improvement in EQCD. A direct, analytical computation is prohibited by sheer complexity, due to the $\OO(a)$-term arising at up to three loops in lattice perturbation theory. A numerical technique was developed to efficiently determine the location of a first order phase transition. This allowed us to obtain the $\OO(a)$ behavior from fitting to a line of constant physics, which was provided by the location of the phase transition at different lattice spacings $\gsqa$. Repeating this calculation at several self-couplings $x$ allowed a grand fit to the parametrically known form of the three-loop counterterm.

Having the expression of the three-loop counterterm in hand, we are now able to study hot QCD free from $\OO(a)$ errors. In particular, a computation of $C(b_\perp)$ and its second moment, $\hat{q}$, is planned. With the improvement available, the leading error in these quantities is down to $\OO(a^2/b_\perp^2)$, which helps especially at small impact parameters $b_\perp$ and makes a continuum limit feasible.

This work was supported by the Deutsche Forschungsgemeinschaft (DFG,
German Research Foundation) -- project number 315477589 -- TRR
211. Calculations for this research were conducted on the Lichtenberg
high performance computer of the TU Darmstadt. We thank Kari Rummukainen, Aleksi Kurkela and Daniel Robaina for useful discussions.

\bibliographystyle{unsrt}
\bibliography{references}

\end{document}